\documentclass[12pt]{article}
\newcommand{\be}[3]{\begin{equation}  \label{#1#2#3}}

\newcommand{\ee}{ \end{equation}}
\newcommand{\ba}{\begin{array}}
\newcommand{\ea}{\end{array}}

\renewcommand{\arraystretch}{1.7}
\let\Large=\large

\begin{document}
\thispagestyle{empty}
\begin{flushright}
HUB-EP-99/22 \\
hep-th/9905180  
\end{flushright}
\vspace{2cm}

\centerline{\bf \Large
Branes, Waves and AdS Orbifolds
}

\vspace{2cm}

\centerline{{\bf Klaus Behrndt}\footnote{Email: behrndt@physik.hu-berlin.de}
\quad and \quad
{\bf Dieter L\"ust}\footnote{Email: luest@physik.hu-berlin.de}
}

\bigskip

\centerline{\em Humboldt-University Berlin, Invalidenstrasse 110,
10115 Berlin, Germany}

\vspace{1cm}

\begin{abstract}
\noindent
The embedding of a Taub-NUT space in the directions transverse
to the world volume of branes
describes branes at (spherical) orbifold singularities.
Similarly, the embedding of a pp-wave 
in the brane world volume 
yields an AdS
orbifold.
In case of the D1-D5--brane system,  
the $AdS_3$ orbifolds yields a BTZ
black hole;  as we will show, the same holds for D3--branes corresponding to
 $AdS_5$.
In addition we will show that the AdS orbifolds and
the spherical orbifolds are U-dual to each other.
However in contrast to spherical orbifolds the AdS orbifolds lead to a
running coupling, which is in the IR inverse to the coupling of the
spherical orbifold.  A discussion of the general pp-wave solution in
AdS space is added.

\end{abstract}

\vspace{5mm}

\newpage

%
%
\section{Introduction}


Besides flat spacetime, the near horizon geometry of regular branes,
given by a direct product of an anti de Sitter (AdS) space and a
sphere, is maximal supersymmetric. By projecting out a discrete
subgroup of the isometry group the supersymmetry will be partially broken. Due
to the direct product of spacetime, there are two distinct possibilities 
\be010
AdS_p \times S_q/\Gamma \qquad ,
\qquad AdS_p/\Gamma \times S_q \ .
\ee

The first class of
orbifolds, called spherical orbifolds,
are well-known from string compactification and
have been discussed in the context of AdS/CFT correspondence \cite{010,
020, 030} in \cite{259, 260, 160}. On the other hand a well-known example
for an $AdS$ orbifold is the BTZ-black hole which corresponds to
$AdS_3/Z_k$ \cite{051}, see also \cite{090}.  While spherical
orbifolds are related to an embedding of a Taub-NUT (TN) space
in the transversal space, we
will show that $AdS$ orbifolds are obtained by (non-standard) pp-wave
embeddings in the brane world volume
directions. Since the anti-de Sitter space is defined as a hyperboloid
in a space with two time-like coordinates, one can embed up to two
waves, i.e.\ one can consider up to two independent orbifolds, which
is also true for spherical orbifolds.  These double orbifolds
correspond to an embedding of intersections of supergravity solutions,
e.g.\ $AdS_5 \times S_5 / (Z_k \times Z_{k'})$ is obtained in the
decoupling limit from the intersection: $D3 \times TN \times TN$ (see
e.g.\ \cite{150, 170}) and as we will argue $AdS_5/(Z_k \times Z_{k'})
\times S_5$ corresponds to the intersection: $D3 \times D(-1) \times
wave$.  Notice, from the F-theory perspective the D(-1)-brane
(D-instanton) is an ``internal'' wave in the two hidden dimensions
\cite{230}.

Obviously, from the supergravity point of view one can dualize the
wave into a Taub-NUT space and vice versa. So, both orbifolds should
be equivalent. 
In fact we will show that the AdS
orbifolds and the spherical orbifolds are
U-dual to each other. But, we have to keep in mind, that both orbifolds acts
very differently: the spherical orbifold breaks the $R$-symmetry,
whereas for the AdS-orbifolds the Lorentz symmetry is broken (see discussion).
The gauge theories, which correspond to the spherical orbifolds, have
reduced number of supersymmetries (compared to $N=4$ in four dimensions),
but are still conformal field theories with vanishing $\beta$-function.
On the other hand,
the AdS-orbifolds yield a running coupling 
from the ultraviolett (UV) to the infrared (IR). 
This issue will be discussed for $AdS_3$ in
the next section and for $AdS_5$ in section 3.
Various aspects of supergravity solutions with broken supersymmetries and
possibly 
running couplings in the corresponding gauge theories, but with unbroken
Lorentz symmetry,  were recently
discussed in \cite{011}-\cite{018}. 


\section{Three-dimensional orbifolds}


The 3-dimensional orbifolds are the simplest cases and the natural framework for
their discussion is the D1-D5--brane system (or also of the F1-NS5),
with the metric given by
\be100
ds^2 = {1 \over \sqrt{H_1 H_5}} \Big( - dt^2 + dy^2 \Big) + \sqrt{H_1 H_5}
        \left( dr^2 + r^2 d\Omega_3  \right) + ds^2_{int} \ .
\ee
For our purpose here, we can identify the two harmonics $H_1 = H_5 = H
= 1 + Q/r^2$ and get a self-dual string in 6 dimensions. In the
decoupling limit (equivalent to the near-horizon limit with large
charges) the internal 4-d metric $ds^2_{int}$ becomes a flat Euclidean
space and the remaining 6-d space becomes $AdS_3 \times S_3$
\be110
ds^2 = \left[{r^2 \over l^2} \Big( - dt^2 + dy^2 \Big) + 
l^2 \, {dr^2 \over r^2} \right]  \ + \ l^2 \,  d \Omega_3   
\quad , \quad l^2 = \sqrt{q_1 q_5} = Q \sim N
\ee
where $l$ is the $S_3$ radius which scales with the number of self-dual
strings $N$.

We start with the discussion of the $S_3$-orbifold ($S_3/Z_k$) and in
order to fix our notation and for later convenience let us repeat some
known facts, see e.g.\ \cite{140}, \cite{080}.  An $S_3$ with a radius
$l$ is defined by
\be130
l^2 = (X^0)^2 + (X^1)^2 + (X^2)^2 + (X^3)^2 = z_1 \bar z_1 + z_2 \bar z_2
\ee
where we introduced the complex coordinates 
\be140
z_1 = l  \, e^{i \, {\theta_R + \theta_L \over 2}} \, 
 \cos {\lambda \over 2} \qquad , \qquad
z_2 = l \, e^{i \, {\theta_R - \theta_L \over 2}}\, 
\sin {\lambda \over 2}  \ .
\ee
Inserting these coordinates, the $S_3$ metric becomes
\be150
ds^2 = dz_1 d\bar z_1 + dz_2 d\bar z_2 
= {l^2 \over 4} 
\Big(d\lambda^2 + d\theta_R^2 + d \theta_L^2 + 2 \cos\lambda \, d\theta_R
d\theta_L \Big) = l^2 d\Omega_3
\ee
where $\lambda \in (0, \pi); \ \theta_R \in (0, 2 \pi) ; \ \theta_L
\in (0, 4\pi)$. As next step we consider the orbifold given by the
lens space $S_3/Z_k$
\be160
z_1 \simeq e^{2 \pi i \over k} z_1 \qquad , \qquad
z_2 \simeq e^{-{2 \pi i \over k}} z_2 
\ee
which corresponds to the identification
\be170
\theta_L \simeq \theta_L + {4 \pi \over k} \ .
\ee
Applying this identification to a given supergravity solution is
equivalent to an embedding of a Taub-NUT-space with NUT charge $p$
($p\sim k$), i.e.
\be180
dr^2 + r^2 d\Omega_3 \ \rightarrow {1\over 1+ {p \over r}} \, 
\Big[ dz + p \, (\pm 1 + \cos\lambda )\, d\phi \Big]^2 + 
(1+{p \over r} ) \Big[dr^2 + r^2 (d\lambda^2 
+ \sin^2 \lambda \, d\theta_R^2 )\Big] \ .
\ee
Making this replacement in (\ref{100}) and taking for the harmonic
functions $H_{1} = H_5 = 1 + {Q \over r}$ and $z = p \, \theta_L$ we
get in the decoupling limit (or near-horizon limit) exactly the same
$S_3$ metric as before, but the consistency of a Taub-NUT space
requires the periodic identification (\ref{170}).

What is the analogous orbifold for the $AdS_3$ space?
As for the sphere, also $AdS_3$ is defined as an hyperboloid,
but now with two time-like directions
\be190
- l^2 = - (X^0)^2 + (X^1)^2 + (X^2)^2 - (X^3)^2 = - z_1^{+} z_1^- + 
z_2^+  z_2^-
\ee
with the four real coordinates given by
\be200
z_1^{\pm} = l  \, e^{\pm \, {\theta_R + \theta_L \over 2}} \, 
 \cosh {\lambda \over 2} \qquad , \qquad
z_2^{\pm} = l \, e^{\pm \, {\theta_R - \theta_L \over 2}}\, 
\sinh {\lambda \over 2}  \ .
\ee
The metric becomes now
\be210
ds^2 = - dz_1^+ dz_1^- + dz_2^+ dz_2^- 
= {l^2 \over 4} 
\Big(d\lambda^2 + d\theta_R^2 + d \theta_L^2 + 2 \cosh\lambda \, d\theta_R
d\theta_L \Big) \ ,
\ee
which coincides with the $AdS_3$ metric given in (\ref{110}) if
$e^{\lambda} =r^2 \gg 1$ and $\theta_{R/L} = y \pm t$. In analogy to
the $S_3$ case the $AdS_3$ orbifold is given by the identification
\be220
z_1^{\pm} \simeq e^{\pm {2 \pi  \over k}} z_1^{\pm} \qquad , \qquad
z_2^{\pm} \simeq e^{\mp {2 \pi  \over k}} z_2^{\pm} 
\ee
which is equivalent to
\be230
\theta_L \simeq \theta_L + {4 \pi \over k}  \ .
\ee
Like the $S_3/Z_k$ orbifold is obtained by an embedding of a Taub-NUT
space in the spherical part of the solution, also the $AdS_3/Z_k$
orbifold corresponds to an embedding of a supergravity solution, but
now it is a pp-wave in the worldvolume part.  In fact, as it has been
widely discussed in matrix theory \cite{220, 200}, (extremal) waves
yields a discretization of a lightcone direction corresponding to a
quantized momentum number.  A wave embedding corresponds to the
replacement
\be240
\ba{c}
-dt^2 + dy^2 \rightarrow -dt^2 + dy^2 + (H-1) (dy - dt)^2  \\
\qquad = {2 \over S_+ + S_-} (dy + S_+ dt)(dy - S_- dt) 
\ea
\ee
with the harmonic function $H = 1 + (r_0/r)^2$ and $S_{\pm} = (H^{-1}
+1) \pm H^{-1}$ ($r_0\sim k$). After this replacement and a shift in the radial
coordinate ($r^2 \rightarrow r^2 - r_0^2$) the $AdS_3$ part in (\ref{110})
becomes
\be242
ds^2 = - \Big({ r^2 - r_0^2 \over r l}\Big)^2 \, 
dt^2 + \Big({ rl \over r^2 - r_0^2 }\Big)^2 dr^2 + \Big({r \over l}\Big)^2
\Big( dy - {r_0^2 \over r^2} dt \Big)^2 \ .
\ee
This is the (extreme) 3-d BTZ black hole which in fact represents an
$AdS_3$ orbifold \cite{051}. It is locally equivalent to (\ref{210})
and globally it is well defined as long as the lightcone direction $y
- t = \theta_L$ in (\ref{240}) is periodically identified.

Therefore, one has the dictionary
\be250
\ba{ccccl}
  D1 \times D5 + TN & \simeq & (AdS_3) & \times & (S_3/Z_k) \\
D1 \times D5 + wave & \simeq & (AdS_3/Z_k) & \times & (S_3)   \ .
\ea
\ee
Of course both configurations are dual to each other, e.g.\ stretching
the branes along the following directions (the NUT and wave directions
are indicated by $\otimes$ and $o$ resp.)
\renewcommand{\arraystretch}{1.1}
\be260
\ba{cccccc}
\hline
1 & 2 & 3 & 4 & 5 & 6 \\ \hline
\times & \times &\times &\times &\times & \\
\times & &&&&  \\
&&&&& \otimes  \\
\hline
\ea
\qquad \leftarrow \quad U \quad
 \rightarrow \qquad
\ba{ccccccc}
\hline
1 & 2 & 3 & 4 & 5 & 6 \\ \hline
 \times &\times &\times &\times &\times & \\
 \times &&&&& \\
 o &&&& & \\ \hline
\ea
\ee
\renewcommand{\arraystretch}{1.7}
the $U$-duality group element, that transforms both configurations
into one another is given by a sequence of $T$ and $S$ dualities
\be270
U =  T_{1236} \; S \; T_{2345} \; S \; T_{1456} \ .  
\ee
Notice, although both orbifolds are duality equivalent, they act very
differently. {From} the worldvolume point of view, the spherical
orbifold ``operates'' in the internal space and leaves the external
spacetime invariant.  On the other hand the AdS orbifold involves the
time or a lightcone direction and corresponds in the decoupling limit
to an infinite momentum frame, as discussed in matrix theory.  In
addition, the periodicity introduced by the orbifold corresponds to a
compact direction with a non-constant radius $R = R(r)$, where $r$ is
the radial coordinate of the AdS space. Effectively this means, that
in contrast to a spherical orbifold, an AdS orbifolds yields a running
couplings.

Let us discuss this point in more detail. Orbifolds change only the
global structure (identifications along a compact direction), locally
the spaces are unchanged i.e.\ they are still $S_3$ or
$AdS_3$. Especially the scalars like the dilaton are not affected by
this procedure, at least not from the supergravity point of view.
However, since the non-trivial global structure corresponds to a
compact coordinate the situation at hand is physically equivalent to
the case where we $T$-dualize this compact direction. By this
T-duality the dilaton will get a dependence on the radius of the
compact direction, i.e.\ the non-trivial global structure becomes
``visible''.  And since the radius of the compact coordinate is not
constant, the dilaton will run from the UV related to $R(r = \infty)$
towards the IR corresponding to $R(r=r_0)$, where the IR point $r=r_0$
is defined as the maximal possible extension while keeping the
coordinate system\footnote{Changing the coordinate system corresponds
in the field theory to an operator reparameterization, which would
imply a different RG behavior \cite{240}.} at $r= \infty$.  Therefore,
the IR appears either as further boundary of spacetime or as a
horizon, see also \cite{100}. To be conrete, let us $T$-dualize the
compact $y$ direction. We obtain from (\ref{242})
\be280
ds^2 = { 1 \over H} \left[ du dv + {r_0^4/l^2 \over r^2}
du^2 \right] + {l^2 \over r^2} \, dr^2 \quad , \quad 
e^{-2 \phi} = {r^2 \over l^2} H \quad , \quad
B= {1 \over 2 H } du \wedge dv
\ee
with $H= 1 + (r_0/r)^2$, and we shifted the horizon from $r= r_0$
in (\ref{242}) to $r=0$ by $r^2 \rightarrow r^2 + r_0^2$.
Asymptotically, this metric  becomes flat  and the dilaton linear
\be290
ds^2 = -dt^2 + dy^2 + l^2 d\lambda^2 \qquad , \qquad
\phi = \lambda
\ee
(${l \over r} = e^{\lambda}$).  Therefore, the dilaton coupling flows
from the UV-free situation towards a non-trivial IR fixpoint near the
horizon
\be300
e^{2\phi} = \left\{ \ba{lll} 0\ , \quad  
 & r\rightarrow \infty & ({\rm UV\ region}) \\
       {l^2 \over r_0^2} \ ,   & r = 0 & ({\rm IR \ region}) \ea \right.
\ee
or if we express the dilaton by dimensionless (integer-valued)
quantities, we find in the IR region
\be310
e^{2\phi_{IR}} \sim {N \over k}
\ee
where $k$ was the momentum number related to the orbifold and $N$ is
the number of self-dual strings.

Let us also compare this result with the $S_3$ orbifold, where we have to
embed the Taub-NUT space in the transversal part.  Because the NUT
direction represents an isometry direction, the D1-D5-brane system is
localized only in 3 transversal coordinates and therefore we have to
replace the harmonic functions by $H_1 = H_5 = 1 +Q/r$.  Hence, making
the replacement (\ref{180}) in (\ref{100}) we find in the decoupling
limit where $Q \sim N$ is large
\be320
\ba{l}
ds^2 = {r \over l^2} \Big( - dt^2 + dy^2 \Big) + 
        {Q \over r} (1 + {p \over r}) \, dr^2  \\ \qquad +
{Q \over r}  {1 \over 1+ p/r} \, \Big[ dz + 
p (\pm 1 + \cos\lambda )\, d\phi \Big]^2 + 
{Q} (r+p)  (d\lambda^2 + \sin^2 \lambda \, d\theta_R^2 )\Big] \ .
\ea
\ee
If in addition $p$ is large the first line becomes the $AdS_3$ space
and the second line represents the $S_3$ orbifold. As for the $AdS_3$
orbifold we can read off an effective coupling from the compact $z$
direction, e.g.\ by employing $T$-duality.  But this time we do not
see a flow behavior for the dilaton. The orbifold itself describes the
field theory in the IR region and the constant dilaton reads
\be330
e^{2\phi_{IR}} = {p \over Q} \sim {k \over N}
\ee
where $k$ is the integer parameterizing the NUT charge $p$, which is
related to the $Z_k$ orbifold ($p \sim k$). Comparing this
expression with (\ref{310}), we see that the coupling constant has
been inverted. The strong-weak coupling duality between both orbifolds
reflects the fact that upon compactification the wave and Taub-NUT
spaces are $S$-dual to each other.


\section{Orbifolds of $AdS_5$}


We will start with the discussion of a single orbifold $AdS_5/Z_k$ and
later we will comment on the double orbifold $AdS_5/(Z_k \times
Z_{k'})$.  In comparison to $AdS_3$, the $AdS_5$ space has
two additional Euclidean coordinates, i.e.\ it is defined
by a hyperboloid in a 6-d space with two timelike directions
\be340
\ba{rcl}
-l^2 &=& -(X^0)^2 + (X^1)^2 + (X^2)^2 - (X^3)^2 + (X^4)^2 + (X^5)^2 \\
&=& - z_1^{+} z_1^- +  z_2^+  z_2^-  + w \bar w
\ea
\ee
with 
\be350
\ba{rcl}
z_1^{\pm} &\equiv & X_0 \pm X_1 = \sqrt{R^2 - l^2} \, e^{\pm \psi_1} \ ,\\
z_2^{\pm}& \equiv & X_2 \pm X_3 = R \, \cos \theta  \, e^{\pm \psi_2} \\
w & \equiv &  X_4 + i X_5 = R \, \sin\theta \, e^{i \chi} \ .
\ea
\ee
The orbifold action on the $AdS_5$ space is defined as before, i.e.
via the embedded
$AdS_3$ space, see eq.(\ref{220}).
Obviously at $\theta = 0$ we get back the $AdS_3$ space as given in
(\ref{190}) with $\psi_{1/2} = { \theta_R \pm \theta_L \over 2}$. In
these coordinates the metric becomes
\be360
ds^2 = - R^2 \cos^2 \theta \, d\psi_2^2 +  (R^2 + l^2) \, d\psi_1^2 +
{l^2 \over R^2 + l^2} \, dR^2 +
R^2 \Big( d\theta^2 + \sin^2\theta d\chi^2 \Big) \ .
\ee
Or, after introducing
\be370
\psi_1 = - {r_+ \over l^2} \, y
\qquad , \qquad 
\psi_2 = { r_+^2 - r_-^2 \over r_+} \, t + {r_- \over l^2} \, y 
\ee
we find
\be380
\ba{l}
ds^2 = \cos^2\theta \, \Big[ - {(r^2 - r_-^2 )(r^2 - r_+^2) \over l^2 r^2} \, 
	dt^2 + {r^2 \over l^2} \Big( dy - {r_- \over r_+} 
	(1 - {r_+^2 \over r^2}) 
	\, dt\Big)^2 \Big] + {l^2 r^2 \over (r^2 -r_-^2 )(r^2 -r_+^2) } \,
	dr^2 \\ 
\qquad + l^2 \, {r^2 -r_+^2 \over r_+^2 - r_-^2} \Big(d\theta^2
	+ \sin^2\theta \, d\chi^2 \Big) + {r_+^2 \over l^2} \, 
	{r^2 -r_-^2 \over r_+^2 - r_-^2}
	\sin^2 \theta \, dy^2 \ .
\ea
\ee
This solution is one example of topological AdS black holes
\cite{070} (see also \cite{060}), which are locally equivalent to the
AdS space, but globally different. In fact, for $\theta = 0$ it
becomes the BTZ black hole with the two horizons at $r = r_{\pm}$.  It
is not only a solution of 5-d Einstein -- anti de Sitter theory, but
it solves also 5-d Chern-Simons with $SO(4,2)$ gauge group; which
includes the Gauss-Bonnet term.

In the extreme limit both horizons coincides and to make the limit
regular, one has also to rescale $\theta$ in a way that
\be390
\theta \rightarrow {\sqrt{r_+^2 - r_-^2} \over l^2 } \, \sigma \quad, \quad
r_+^2 \rightarrow r_-^2 = r_0^2 \qquad {\rm with}: \qquad \sigma \ {\rm fix.}
\ee
After replacing $r^2 - r_0^2 \rightarrow r^2$,  $2t = v$, $y =u$ we 
obtain a 3-brane with a pp-wave 
\be400
ds^2 = {r^2 \over l^2} \Big[ -dv \, du + H du^2 + d\sigma^2 + 
	\sigma^2 d\chi^2
	\Big] + l^2 \, {dr^2 \over r^2}
\quad , \qquad
H= 1 + {r_0^2 \over l^4} \, \sigma^2 + {r_0^2 \over r^2} \ .
\ee
but due to the $\sigma$ dependence of $H$ it is {\em not} the standard wave.  

To understand this solution better let us discsuss the general wave
solution in an AdS space, see also \cite{120}.  The standard wave
ansatz in an $AdS_{p+2}$ space reads ($z = {l^2 \over r}$)
\be410
ds^2 = {l^2 \over z^2} \Big[ -dv du + H(u,x_i, z) \, 
	du^2 + dx_1^2 + \cdots + dx_{p-2}^2 + dz^2 \Big] \ .
\ee
This ansatz is a solution of $AdS_{p+2}$ gravity ($R_{\mu}^{\ \nu} = -{p+1
\over l^2} \delta_{\mu}^{\ \nu}$) if $H$  solves the Laplace
equation
\be420
\Delta H \equiv {1 \over \sqrt{g} } \partial_{\mu} \Big(\sqrt{g}
g^{\mu\nu} \partial_{\nu} H \Big) = 
{z^2 \over l^2} \Big( \partial_{\|}^2 + z^{p} \partial_z \, z^{-p} 
\partial_z \Big) H(u, \vec x , z) = 0
\ee
where $\vec x = (x_1, x_2, \ldots )$ and, as typical for pp-waves, one
can allow for a general $u$ dependence; for a recent discussion of
this $u$ dependence in the AdS/CFT correspondence see e.g.\ \cite{280}.  
Let us give some special cases:

(i) if $H = H(z)$: the solution is $H= h + c z^{p+1}$,
which corresponds to the expected harmonic function with respect to
the $9-p$ transversal brane directions (note $z \sim 1/r$).  
\newline
(ii) if $H = H(\vec x)$: one gets an harmonic function with respect to
the worldvolume coordinates, e.g.\ a logarithm for the 3-brane.
\newline
(iii) if $H = f(\vec x) g(z)$: the solution becomes $H \sim e^{\pm i \vec p
\cdot \vec x} \, z^{\nu} {\cal K}_{\nu} (|p| z)$, $\nu = {1 + p \over
2}$ where ${\cal K}_{\nu} (|p| z)$ is the modified Bessel functions.
At the asymptotic boundary ($z=0$, UV region) $z^{\nu} {\cal K}_{\nu}
\sim constant$ and $H$ parameterize a plane wave 
whereas in the IR ($z \rightarrow \infty$) $H$ vanishes.
\newline
(iv) finally if $H = H(|\vec x|^2 + z^2)$: the solution reads
\be430
H = h + c (|\vec x|^2 + z^2) \ .
\ee
For $AdS_5$ this last solution coincides exactly with solution obtained in
(\ref{400}) if $|\vec x| = \sigma$; $z = l^2/r$ and $c =
r_0^2/l^4$. Notice, that this regular solution is the same exact wave
solution which is also known from the flat space case. Moreover, this
wave breaks explicitly the world volume isometries; only the lightcone
direction $u, v$ represents still isometries.

By a simple shift in $v$ we can absorb $h$ and in order to avoid a
conical singularity at $\sigma = z = 0$ we have to do a periodical
identification along the lightcone direction $u$, which corresponds
exactly to our orbifold. This was also the case for $AdS_3$: for
the non-extreme case the identification is along a spatial direction
whereas in the extreme case a lightcone direction is periodically
identified. Moreover, this compact direction breaks the scale
invariance and since its radius is not constant, it corresponds to a
running coupling. Again reading off the dilaton from this compact
direction we find $e^{-2 \phi} = g_{uu} = {r^2 \over l^2} H$ or
\be450
e^{2\phi} = \left\{ \ba{lll} 0 \ ,\quad  
 & r\rightarrow \infty & ({\rm UV\ region}) \\
       {l^2 \over r^2_0}
	 \sim {N \over k}\ , & r = 0 & ({\rm IR \ region}) \ea \right.
\ee
Notice, that the $\sigma$-dependence drops out!

We can also compare this result with the $S_5$ orbifold, related to a
Taub-NUT embedding into the $S_5$. Reading off the dilaton coupling
from the radius of the NUT direction, we find
\be470
e^{2\phi_{IR}} \sim {k \over N}
\ee
where the radius of $S_5$ scales with $N$ and $k$ corresponds to the
NUT charge which corresponds to the orbifold. Recall for this
spherical orbifold we do not get a running coupling, but only the
effective value in the infrared (the compact NUT direction introduces
a scale), which is again opposite to the value from the $AdS$
orbifold.

Finally, let us also comment on the double orbifolds $AdS_5/(Z_k
\times Z_{k'})$. Because both orbifolds appear in a democratic way,
also the second $Z_{k'}$ factor corresponds to an embedding of a
pp-wave. Because $AdS_5$ was defined as a hyperboloid in the a 6-d
space with {\em two} time-like direction, this space can accommodate
exactly two independent waves (a wave requires always a time-like
direction). This agrees nicely with the spherical orbifolds, which
also allow for at most two independent orbifold actions. But reducing
the 6-d space to the $AdS_5$ we loose one time and thus the second
wave cannot show up as a second wave in $AdS_5$. On the other hand
from the F-theory approach to type IIB string theory, we know that the
$D$-instanton corresponds exactly to a wave with respect to the
``hidden time'' \cite{230}. Hence, the corresponding supergravity
solution has to include a wave as well as a $D$-instanton and it is
straightforward to construct this solution.  First note, that the
3-brane metric (\ref{410}) is the same in the Einstein and string frame
and therefore we can interprete it as the Einstein metric. In the
Einstein frame the IIB dilaton $e^{-\phi}$ and RR scalar $l$ 
($S^\pm=l\pm e^{-\phi}$)
solve the
equations of motion if (see e.g.\ \cite{250})
\be480
S^+={\rm const.},\quad S^-=
{1 \over \bar H},
\ee
where $\bar H$ is also a solution of the Laplace equation (\ref{420}).
Thus in the string frame we obtain the metric
\be490
ds^2 = \sqrt{\bar H} \, {l^2 \over z^2} \Big[ -dv du + H(u,x_i, z) \, 
	du^2 + dx_1^2 + dx_2^2 + dz^2 \Big]  \ .
\ee
The harmonic function $\bar H$ could be any the cases discussed after
(\ref{420}). For further discussions of $D$-instantons in the
AdS/CFT correspondence we refer to \cite{110, 270, 277}


\section{Discussion}


In the near-horizon limit, regular branes factorize into an anti de
Sitter space and a sphere yielding two distinct possibilities for
orbifolds: $AdS \times S/\Gamma$ or $AdS/\Gamma \times S$.  In this
paper we focused on AdS-orbifolds and compared them with spherical
orbifolds. Both cases correspond to an embedding of a supergravity
solution: a Taub-NUT space for spherical orbifolds and  pp-waves for
the AdS-orbifolds.  Both the supergravity solutions acts quite
differently, the Taub-NUT space breaks the $R$-symmetry related to
rotations in the transverse space and the wave breaks the worldvolume
rotational symmetry.  {From} the field theory point of view the
$R$-symmetry is internal, whereas the worldvolume rotations broken by
pp-waves are part of the Lorentz group. Notice, pp-waves are
interpreted as gravitons with a given momentum and in the decoupling
limit this momentum should be large.

In projecting out the discrete subgroup, one has to truncate the
spectrum onto an invariant subsector. For both orbifolds one has to
project out one chirality: for spherical orbifolds it is one chirality
with respect to the NUT direction, as discussed in \cite{140} for the
3-d case; and for the AdS-orbifold it is one chirality with respect to the
momentum modes corresponding to the wave. For the $AdS_3$ case the
corresponding states are the chiral primaries corresponding to
momentum modes travelling only in one direction along the (extremal)
string, see \cite{040}.

Finally, let us stress that the pp-wave yielding an AdS-orbifold is in
general not the ``standard'' pp-wave that corresponds to a harmonic
function. Instead, it is the exact wave solution that quadratically
increases in the transversal space, see (\ref{430}), and the periodic
identification, which avoids a conical singularity, represents the
orbifold.  So, this solution breaks the translational
invariance along the brane worldvolume (localized wave). Only the wave
direction represents still an isometry direction and by T-dualizing
this direction one can construct brane intersection which are
localized and supersymmetric.

\vskip0.5cm
\noindent{\bf Acknowledgements:}

\noindent 
Work partially supported by the Deutsche Forschungsgemeinschaft (DFG) and
by the E.C. project
ERBFMRXCT960090.


%
%
\providecommand{\href}[2]{#2}\begingroup\raggedright\endgroup


\end{document}